\documentstyle[aps,preprint]{revtex}
\begin{document}
\draft
\tighten
\title{The Georgi ``Avatar'' of Broken Chiral Symmetry in Quantum Chromodynamics}
\author{R. Acharya\footnote{Electronic address: raghunath.acharya@asu.edu}}
\address{Physics Department,
Arizona State University, Tempe, AZ 85287}
\author{P. Narayana Swamy\footnote{Electronic address: pswamy@siue.edu}}
\address{Physics Department, Southern
Illinois University, Edwardsville, IL 62026}
\date{\today}
\maketitle
\begin{abstract}
We establish that in Quantum Chromodynamics (QCD) at zero
temperature, $SU_{L+R}(N_F)$ exhibits the vector mode conjectured
by Georgi and  $SU_{L-R}(N_F)$ is realized in either the
Nambu-Goldstone mode or else $Q_5^a$ is also screened from view at
infinity. The Wigner-Weyl mode is ruled out unless the beta
function in QCD develops an infrared stable zero.
\end{abstract}

\vskip 4ex
PACS numbers: \pacs{11.3 Qc., 11.30.Rd., 12.38.Aw}
\pagebreak

We shall establish that QCD at zero temperature, satisfying both
asymptotic freedom and confinement, exhibits the following
features: (a) $SU_{L+R}(N_F)$ exhibits the vector mode conjectured
by Georgi \cite{bib:georgi} (Georgi-Goldstone mode). (b)
$SU_{L-R}(N_F)$ exhibits either the Nambu-Goldstone mode or else
the axial-vector charge $Q_5^a$ is also screened from view at
infinity. If the latter case were to occur, then QCD confines
without breaking chiral symmetry: both $SU_{L+R}(N_F)$ and
$SU_{L-R}(N_F)$are realized in the Higgs mode (Georgi-Wigner
mode), with no scalar or pseudoscalar Nambu-Goldstone bosons and
the vector and axial-vector mesons are degenerate. (c) The
Wigner-Weyl mode corresponding to $Q^a \left |\: 0 \left \rangle
\right. \right. = 0$, $Q_5^a \left |\: 0 \left \rangle
\right. \right. = 0$ is ruled out: the Callan-Symanzik beta
function has to turn over to yield an infrared stable fixed point
at a finite value of g if chiral symmetry is to be restored.

We now sketch the proof of these interesting assertions. We begin
with the vector current $V_\mu ^a$ and its conservation
\begin{equation}
\partial^\mu V_\mu ^a \left( {\bf x},t \right) =0.
\end{equation}
This implies the local version
\begin{equation}
\left[ Q^a \left(t \right), H \left( {\bf x},t \right) \right] = 0
\label{eq:two}
\end{equation}
where $ H \left( {\bf x},t \right)=\Theta^{00}$ is the Hamiltonian
density, if the surface terms at infinity can be discarded. This
is clearly justified if the flavor vector charge annihilates the
vacuum,
\begin{equation}
Q^a(t) \left |\: 0 \left \rangle \right. \right. = 0
\end{equation}
which is guaranteed by the Vafa-Witten theorem
\cite{bib:vafa,bib:acharya89} {\it i.e.,} non-chiral symmetries
cannot be spontaneously broken in vector-like gauge theory. Hence
there are no scalar Nambu-Goldstone bosons to produce a long range
interaction, which in turn would have resulted in a non-vanishing
contribution to the surface terms.

The dilatation charge
\begin{equation}
Q_D(t) = \int d^3x D_0\left({\bf x},t\right),
\label{eq:four}
\end{equation}
defined in terms of the dilatation current $D_\mu \left( {\bf x},t
\right)$ satisfies \cite{bib:adler} the trace anomaly
\begin{equation}
\partial ^\mu D_\mu = \frac{\beta(g)}{2g} G_{\mu
\nu}^{\alpha} G_\alpha^{\mu \nu}
\label{eq:five}
\end{equation}
in QCD in the chiral limit when the current quark mass is zero. It
is well-known that scale invariance is broken both
``spontaneously'', $Q_D(t) \left |\: 0 \left \rangle \right. \right. \ne
0$, and explicitly by the trace anomaly. Consequently, the states
defined by successive repeated application of $Q_D(t)$ on the
vacuum state are neither vacuum states nor are they necessarily
degenerate \cite{bib:archarya97}. Let the commutator
\begin{equation}
\left[Q_D(0),Q^a(0) \right] = -id_Q Q^a(0)
\label{eq:six}
\end{equation}
define the scale dimension $d_Q$ of the charge $Q^a(0)$. By
translation invariance, this can be put in the form
\begin{equation}
\left[Q_D(t),Q_a(t) \right] = -id_Q Q_a(t). \label{eq:seven}
\end{equation}
It is important to stress that operator relations such as the
above equation are unaffected by spontaneous symmetry breaking as
emphasized by Weinberg \cite{bib:weinberg}. Let us consider the
double commutator which follows from Eq.\ (\ref{eq:two}),
\begin{equation}
\left[ Q_D(t), \left[Q_a(t), H \right] \right] =0,
\end{equation}
where $Q_D(t)$ is the dilation charge defined in Eq.\ (\ref{eq:four}).
If we now invoke the Jacobi identity we can
recast the above equation in the form
\begin{equation}
\left[Q_a(t),\left[H,Q_D(t)\right]\right] +
\left[H,\left[Q_D(t),Q_a(t)\right] \right] =0.
\end{equation}
Since
\begin{equation}
\left[ H({\bf x},t),Q_D(t) \right] = -i \partial_\mu D^\mu ({\bf
x},t) \ne 0,
\end{equation}
by virtue of the trace anomaly, Eq.\ (\ref{eq:five}), and making
use of Eqs.\ (\ref{eq:two},\ref{eq:seven}), we arrive at the
operator relation
\begin{equation}
\left[Q_a(t),\partial ^\mu D_\mu({\bf x},t) \right] =0.
\label{eq:eleven}
\end{equation}
Applying this relation on the vacuum state, we obtain
\begin{equation}
\left[Q_a(t),\partial ^\mu D_\mu({\bf x},t) \right] \left | \: 0 \left \rangle \right. \right. =0.
\label{eq:twelve}
\end{equation}
We may now invoke the result of Vafa-Witten theorem
\cite{bib:vafa}
\begin{equation}
Q_a(t) \left | \: 0 \left \rangle \right. \right. =0,
\end{equation}
and conclude that
\begin{equation}
{\cal O}({\bf x},t) \left | \: 0 \left \rangle \right. \right.
\equiv Q_a(t) \partial ^\mu D_\mu({\bf x},t) \left | \: 0 \left
\rangle \right. \right. =0,
\label{eq:fourteen}
\end{equation}
where the operator ${\cal O}({\bf x},t)$ is {\it local} in space
and time and commutes with itself for space-like intervals.
Explicitly, we see that this condition reduces to
\begin{equation}
\left[ Q^a(t) \partial ^\mu D_\mu ( {\bf x},t), Q^a(t^\prime)
\partial ^\mu D_\mu ( {\bf x}^\prime,t^\prime)\right] =
Q^a(t)Q^a(t^\prime) \left[ \partial ^\mu D_\mu({\bf x},t),\partial
^\mu D_\mu({\bf x}^\prime,t^\prime) \right]=0 \label{eq:fifteen}
\end{equation}
for space-like intervals, due to the time independence of $Q^a(t)$
and Eq.\ (\ref{eq:eleven}). Here we have invoked the locality of
$\partial^\mu D_\mu({\bf x},t)$.

We can now utilize the Federbush-Johnson theorem
\cite{bib:coleman}, which applies to any local operator, to Eq.\
(\ref{eq:fourteen}) and immediately arrive at the key result,
\begin{equation}
Q_a(t) \partial ^\mu D_\mu ({\bf x},t) \equiv 0.
\label{eq:sixteen}
\end{equation}
Since $\partial^\mu D_\mu ({\bf x},t)$ cannot vanish in a theory
which exhibits both asymptotic freedom and confinement except at
$g=0$, we conclude that the vector flavor charges must be screened
from view at infinity,
\begin{equation}
Q^a(t)=0.
\end{equation}
This important result is a manifestation of spontaneously broken
local symmetry. The vector mesons become massive and the scalar
would-be Nambu-Goldstone bosons disappear.

Let us now consider the axial-vector charges $Q_5^a$. The
Vafa-Witten theorem does not apply in this case and therefore we
proceed by the method of {\it reductio ad absurdum} as follows. We
assume $Q_5^a \left| \: 0 \left \rangle \right. \right.=0$
corresponding to the Wigner-Weyl mode of unbroken symmetry. We
begin by defining the scale dimension of the axial-vector charge
by
\begin{equation}
\left[Q_D(0),Q_5^a(0) \right] = -id_{Q _5}Q_5^a(0).
\end{equation}
Repeating the earlier analysis now for the axial-vector charges,
exactly as in Eqs.\ (\ref{eq:seven}--\ref{eq:fifteen}), we arrive
at the conclusion
\begin{equation}
Q_5^a(t)\partial ^\mu D_\mu ({\bf x},t) \equiv 0.
\label{eq:nineteen}
\end{equation}
Since $Q_5^a(t) \ne 0$ for the assumed Wigner-Weyl mode, this
requires $\partial^\mu D_\mu ({\bf x},t) =0$. This would imply
that QCD exhibiting both asymptotic freedom and confinement is
free. Hence by {\it reductio ad absurdum} we are led to the
conclusion: either $Q_5^a(t) \left | \: 0 \left \rangle \right.
\right. \ne 0$ or $Q_5^a(t) \equiv 0$. The first alternative is
the Nambu-Goldstone realization of chiral symmetry which must hold
for  $N_F=3$, \cite{bib:weinberg}. The second alternative in
conjunction with the screening of the vector charges, {\it i.e.},
$Q^a(t)\equiv 0$, $Q_5^a(t)\equiv 0$, is the Higgs mode
alternative (Georgi-Wigner mode): both scalar and pseudoscalar
Nambu-Goldstone bosons have been devoured. This case corresponds
to confinement with exact chiral symmetry. Such a mode is realized
in Supersymmetric QCD \cite{bib:intriligator}.

In conclusion, it is interesting that the Wigner-Weyl mode is
ruled out in QCD at $T=0$, if both asymptotic freedom and
confinement obtains. this follows from Eqs.\
(\ref{eq:sixteen},\ref{eq:nineteen}). Hence the Wigner-Weyl mode
can occur only if the beta function turns over to yield an
infrared stable fixed point \cite{bib:appelquist}.  Which of the
above alternative realizations indeed occurs in QCD cannot be
settled by this analysis. The answer may depend on the number of
flavors.

An effective Lagrangian \cite{bib:bando,bib:appelquist1} which
realizes the Georgi-Goldstone mode is easily constructed,
following the genral procedure for building models with vector and
axial vector mesons.

 Finally, it is important to stress that we have not
invoke the $SU_L(N_F) \times SU_R(N_F)$ charge algebra in our
analysis. In view of the screened charges, {\it i.e.},
$Q^{a}(t)=0$ for the Georgi-Goldstone mode and $Q^a(t)=Q_5^a(t)=0$
for the Georgi-Wigner mode, one must revert back to current
algebra \cite{bib:weinberg} which leads to the Weinberg sum rules
in QCD.

\acknowledgments

We are grateful to Professor S. Adler for drawing our attention to
the work of Intriligator and Seiberg, and Seiberg and Witten; in
particular to the phase of confinement without chiral breaking in
supersymmetric QCD for $N_F= N_C+1$. The title of the paper is
inspired by a famous paper entitled ``Fermion avatars of the
Sugawara model'' by S. Coleman {\it et.\ al.} in mid-1970's.

\end{document}